\documentclass[runningheads]{llncs}
\usepackage[T1]{fontenc}
\usepackage{booktabs}
\usepackage{graphicx,verbatim}
\usepackage[numbers]{natbib}
\usepackage{hyperref}
\usepackage{xcolor}
\usepackage{amsmath}
\usepackage{amsfonts}
\usepackage[colorinlistoftodos]{todonotes}
\usepackage{threeparttable}
\usepackage{subcaption}

\begin{document}
\title{Analysis of Transferability Estimation Metrics for Surgical Phase Recognition}
%
\author{Prabhant Singh\inst{1}\orcidID{0000-0001-7634-2312} \and
Yiping Li\inst{2,3}\orcidID{0009-0005-0239-3682} \and
Yasmina Al Khalil\inst{3}\orcidID{0000-0001-6839-3507}}
\institute{Eindhoven University of Technology \\
     \email{\{p.singh, y.li9, y.al.khalil\}@tue.nl}}

\maketitle              
\begin{abstract}
Fine-tuning pre-trained models has become a cornerstone of modern machine learning, allowing practitioners to achieve high performance with limited labeled data. In surgical video analysis, where expert annotations are especially time-consuming and costly, identifying the most suitable pre-trained model for a downstream task is both critical and challenging. Source-independent transferability estimation (SITE) offers a solution by predicting how well a model will fine-tune on target data using only its embeddings or outputs, without requiring full retraining. In this work, we formalize SITE for surgical phase recognition and provide the first comprehensive benchmark of three representative metrics, LogME, H-Score, and TransRate, on two diverse datasets (RAMIE and AutoLaparo). Our results show that LogME, particularly when aggregated by the minimum per-subset score, aligns most closely with fine-tuning accuracy; H-Score yields only weak predictive power; and TransRate often inverses true model rankings. Ablation studies show that when candidate models have similar performances, transferability estimates lose discriminative power, emphasizing the importance of maintaining model diversity or using additional validation. We conclude with practical guidelines for model selection and outline future directions toward domain-specific metrics, theoretical foundations, and interactive benchmarking tools.\footnote{Accepted at 3rd Workshop in Data Engineering in Medical Imaging (DEMI), MICCAI-2025 Workshop}

\end{abstract}
\section{Introduction}

In recent years, deep learning (DL) has played a central role in advancements in surgical artificial intelligence (AI), particularly in the domain of surgical video understanding . Among these, surgical phase recognition, the task of classifying each frame of a surgical video into predefined procedural stages, has received significant attention due to its potential to support intra-operative guidance, post-operative analytics, and autonomous robotic assistance. However, despite promising results, the widespread adoption of DL models in surgical AI remains hindered by a persistent challenge: the lack of annotated surgical video data.

Acquiring annotations for surgical videos is a highly resource-intensive process, requiring domain expertise, time, and effort. Moreover, the surgical environment presents substantial variability and complexity: frequent occlusions, motion blur, illumination changes, and heterogeneous surgical styles across institutions and surgeons. These factors, coupled with the high temporal and contextual dependencies inherent in surgical workflows, make it difficult to train generalizable and reliable DL models without large-scale, high-quality labeled datasets. To address data scarcity, transfer learning has become a widely adopted strategy. Models pre-trained on large-scale datasets such as ImageNet\cite{imagenet} are often fine-tuned on specific medical imaging tasks, including surgical phase recognition. This approach can significantly boost performance and reduce training time, especially in low-data regimes. Recent works have demonstrated that large-scale pre-training improves downstream performance in surgical tasks \cite{jaspers2025scaling, batic2024endovit, schmidgall2024general}. However, the effectiveness of transfer learning in surgical applications remains uncertain. The domain gap between natural images and surgical data often limits the utility of off-the-shelf pre-trained models, and the performance gain is heavily dependent on factors such as model architecture, source dataset characteristics, and fine-tuning strategies.

Given the rapidly growing landscape of available pre-trained models, selecting the most suitable model for a specific surgical application has become a non-trivial problem. Exhaustive fine-tuning of each candidate model is computationally expensive and impractical, particularly in time- and resource-constrained clinical settings. As such, there is a critical need for transferability estimation methods that can predict how well a pre-trained model is likely to perform on a surgical downstream task without extensive retraining. This challenge, known as source-independent transferability estimation (SITE), seeks to evaluate model suitability without relying on fine-tuning or access to source training data. While SITE has been explored in natural image classification and general domain adaptation contexts, its application to surgical video understanding remains underexplored. Effective SITE methods could drastically reduce the cost of model selection and adaptation in surgical AI, enabling more efficient use of scarce annotated data and computational resources.

In this work, we frame Surgical Phase Recognition as a testbed for investigating SITE in the context of surgical AI. We highlight the limitations of current pre-training paradigms for surgical tasks and motivate the need for principled transferability assessment approaches tailored to the unique challenges of the surgical domain. Our code for the experiments can be found on \url{https://github.com/prabhant/surgical_phase_transferability_estimation}



\subsection{Related Work}
In the context of Source-Independent Transferability Estimation, numerous metrics have been proposed, each grounded in different theoretical perspectives and methodological assumptions.  LogME~\cite{logme} formalizes the transferability
estimation as the maximum label marginalized likelihood
and adopts a directed graphical model to solve it. SFDA~\cite{SFDA} proposes a self-challenging mechanism, it first maps the features and then calculates the sum of log-likelihood as the metric. NCTI~\cite{NCTI} treats it as a nearest centroid classifier problem and measures how close the geometry of the target features is to their hypothetical state in the
terminal stage of the fine-tuned model. LEEP~\cite{nguyen2020leep} is the average log-likelihood of the log-expected empirical predictor, which is a non-parameter classifier based on the joint distribution of the source and target distribution, N-LEEP~\cite{NLEEP} is a further improvement on LEEP by substituting the output layer with a Gaussian
mixture model. TransRate~\cite{transrate2} treats SITE from an information theory point of view by measuring the transferability as the mutual information between features of target examples extracted by a pre-trained model and their labels.

In the context of SITE in medical imaging, only a limited number of studies have explored transferability estimation. Notably, Juodelyte et al.\cite{juodelyte2024datasettransferabilitymedicalimage} introduced a dynamic nearest component analysis-based metric for medical image classification in a comprehensive evaluation setup, while CC-FV\cite{segtransfer} was proposed for SITE in the domain of medical image segmentation.




\section{Methodology}
In this section, we (i) formulate the general transferability estimation problem (\ref{sec:problem_statement}), (ii) tailor the framework to the surgical phase recognition task (\ref{sec:problem_statement_phase}), and (iii) outline our experimental setup for assessing the effectiveness of different transferability metrics (\ref{sec:experiment}). 

\subsection{Problem Statement}
\label{sec:problem_statement}

We assume that we are given a target dataset $\mathcal{D} = \{(\mathbf{x}_n,y_n)\}_{n=1}^N$ of $N$ labeled points and $M$ pre-trained models $\{\Phi_m=(\phi_m, \psi_m)\}_{m=1}^M$. Each model $\Phi_m$ consists of a feature extractor that returns a $d$-dimensional embedding $\phi_m(x)\in\mathbb{R}^d$ and the final layer or head $\psi_m$ that outputs the label prediction for the given input $x$. \begin{figure}
    \centering
    \includegraphics[width=\linewidth]{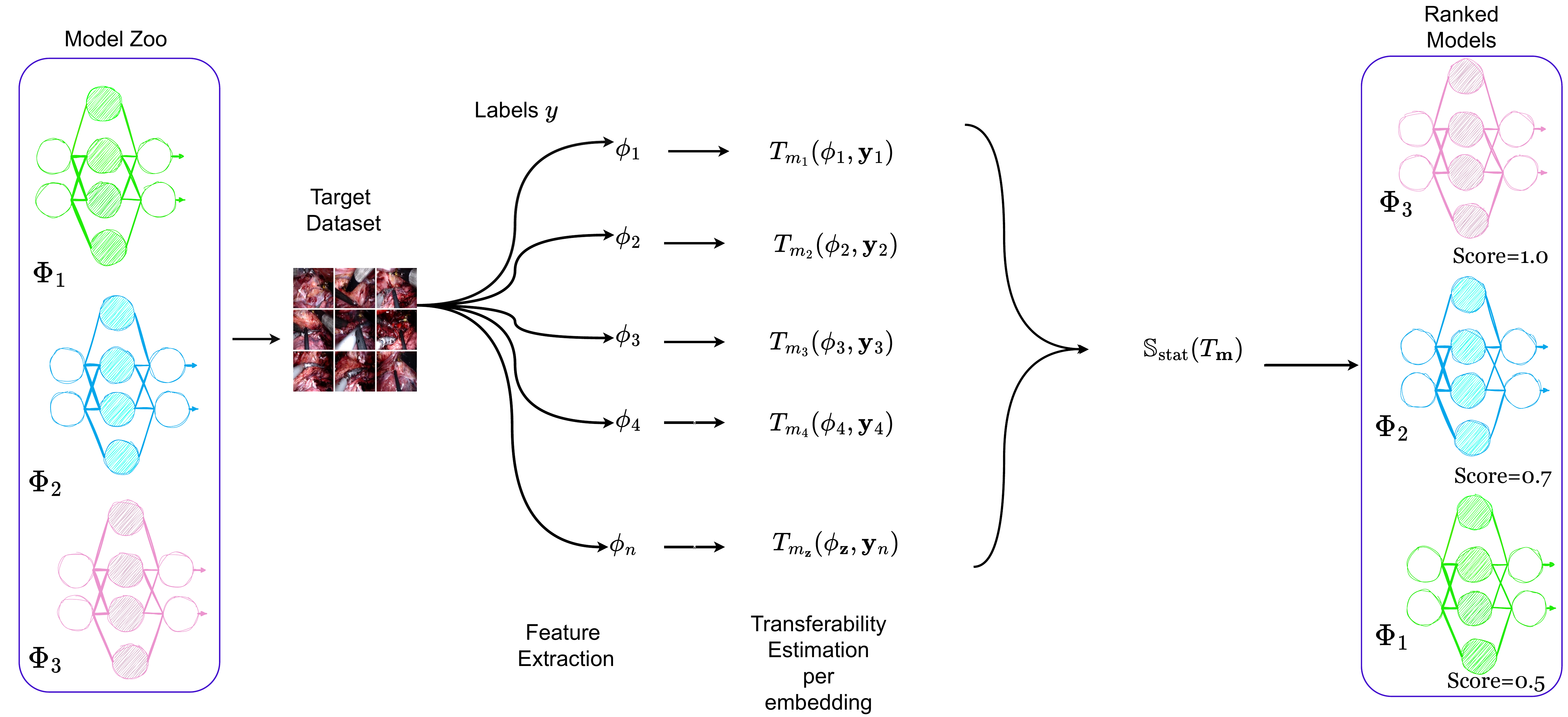}
    \caption{Estimating transferability via frame-wise estimation.}
    \label{fig: probstatement}
\end{figure}The task of estimating transferability is to generate a score for each pre-trained model so that the best model can be identified via a ranking list. For each pre-trained model $\Phi_m$ a transferability metric outputs a scalar score $T_m$ that should be coherent in its ranking with the performance of the fine-tuned classifier $\hat{\Phi}_m$. That is, the goal is to obtain scores $T_m$ as:  \\
\begin{equation}
     T_m \geq T_n \Leftrightarrow  \frac{1}{N}\sum_{n=1}^Np(y_n|\mathbf{x}_n; \hat{\Phi}_m) \geq \frac{1}{N}\sum_{n=1}^Np(y_n|\mathbf{x}_n; \hat{\Phi}_n),
\end{equation}
\\
where $p(y_n|x_n; \hat{\Phi}_m)$ indicates the probability that the fine-tuned model $\hat{\Phi}_m$ predicts label $y_n$ for input $\mathbf{x}_n$.
A larger $T_m$ indicates better performance model on target data $\mathcal{D}$.

\subsection{Application to Surgical Phase Recognition}
\label{sec:problem_statement_phase}
In the specific context of \textit{surgical phase recognition}, the target dataset $\mathcal{D}$ is composed of $\mathbf{z}$ sampled static frames taken from surgical videos, with each frame annotated by a phase label. These frames may be grouped into $Z$ disjoint subsets (e.g., surgeries or temporal segments), each consisting of $N$ labeled examples.

The dataset can thus be expressed as:

\[
\mathcal{D} = \bigcup_{a=1}^{Z} \left\{(\mathbf{x}_n^{(a)}, y_n^{(a)})\right\}_{n=1}^{N}
\]

For a given pre-trained model $\Phi_m = (\phi_m, \psi_m)$, we extract feature embeddings for all frames in subset $a$ using $\phi_m$ and compute a transferability score $T_m^{(a)}$ based on the extracted features and corresponding labels:

\[
T_m^{(a)} = T_m(\phi_m(\mathbf{X}^{(a)}), \mathbf{y}^{(a)}),
\quad \text{where } \mathbf{X}^{(a)} = \{\mathbf{x}_n^{(a)}\}_{n=1}^{N}, \quad \mathbf{y}^{(a)} = \{y_n^{(a)}\}_{n=1}^{N}
\]

We collect all such scores across the $Z$ subsets:

\[
T_{\mathbf{m}} = \{T_m^{(1)}, T_m^{(2)}, \ldots, T_m^{(Z)}\}
\]

To obtain a global transferability score for model $\Phi_m$, we apply a statistical summarization function $\mathbb{S}_{\text{stat}}$:

\[
T_m = \mathbb{S}_{\text{stat}}(T_{\mathbf{m}})
\]

Here, $\mathbb{S}_{\text{stat}}$ may be a mean, median, or any robust summary statistic. In our experiments we instantiate $\mathbb{S}_{\text{stat}}$ as three different summary functions: mean, minimum, and maximum, so as to explore how outlier‐sensitive vs. robust summaries affect the global transferability ranking. This global score $T_m$ is then used to rank all models, with higher values indicating a greater potential for successful fine-tuning in the surgical phase recognition task.

The goal remains to ensure that these scores correlate well with the actual performance of the fine-tuned models $\{\hat{\Phi}_m\}_{m=1}^M$ on the surgical dataset. Figure \ref{fig: probstatement} illustrates our transferability‐estimation pipeline.

\subsection{Experimental setup}
\label{sec:experiment}
\textbf{Data:} In this study, we investigate source-independent transferability for surgical phase recognition using two datasets. The first is the in-house RAMIE dataset \cite{10.1117/12.3040084}, which contains 27 thoracoscopic RAMIE recordings annotated with 13 distinct surgical phases. 
We sample all videos at 1 fps and resize frames to 256×256 pixels, yielding 40,881 frames for training (14 videos), 27,249 for validation (4 videos), and 65,596 for testing (9 videos). Phase labels were established by two expert annotators with a consensus-driven review. The second dataset is the publicly available AutoLaparo dataset \cite{wang2022autolaparo}, comprising full-length videos of complete hysterectomy procedures annotated with seven surgical phases. After identical 1 fps sampling and resizing, it is split into 40,211 frames (10 videos) for training, 12,056 frames (4 videos) for validation, and 12,056 frames (9 videos) for testing. These datasets cover distinct procedures and annotation schemes, enabling assessment of transferability methods under varied conditions.\\
\textbf{Fine-Tuning details:} We generate frame-wise features by applying the pre-trained models listed in Table \ref{tab:model_overview} to every frame of the training and validation videos in both datasets.
\begin{table}[h!]
\centering
\caption{Overview of model architectures, parameters, and pre-training datasets}
\begin{tabular}{@{}l l c l@{}}
\toprule
\textbf{Model Name}      & \textbf{Backbone}   & \textbf{Param} & \textbf{Pre-training Dataset} \\
\midrule
MetaFormer \cite{yu2023metaformer}               & CAFormer \cite{yu2023metaformer} & 27.5M                 & ImageNet \\
MetaFormer\_in21k \cite{yu2023metaformer}        & CAFormer \cite{yu2023metaformer}            & 27.5M                 & ImageNet21k \\
SurgeNet\_Cholec 
 \cite{jaspers2025scaling}          & CAFormer \cite{yu2023metaformer}            & 27.5M                 & Cholecystectomy collection \\
SurgeNet\_Public \cite{jaspers2025scaling}           & CAFormer \cite{yu2023metaformer}            & 27.5M                 & Public surgical videos \\
SurgeNet\_XL \cite{jaspers2025scaling}               & CAFormer \cite{yu2023metaformer}            & 27.5M                 & Public \& inhouse surgical videos \\
TimeSformer \cite{bertasius2021space}              & ViT Base \cite{dosovitskiy2020image}             & 121.4M                  & Kinetics-400 dataset \\
ConvNeXtV2 \cite{liu2022convnet}               & ConvNeXtV2 \cite{liu2022convnet}          & 31.9M                 & ImageNet \\
SurgeNet\_ConvNeXtV2 \cite{jaspers2025scaling}       & ConvNeXtV2 \cite{liu2022convnet}          & 31.9M                 & Public \& inhouse surgical videos \\
PVTv2 \cite{wang2021pyramid}                    & PVTv2 \cite{wang2021pyramid}               & 28.0M                 & ImageNet \\
SurgeNet\_PVTv2 \cite{jaspers2025scaling}            & PVTv2 \cite{wang2021pyramid}               & 28.0M                 & Public \& inhouse surgical videos \\
GastroNet\_RN50 \cite{boers2024foundation}          & ResNet-50 \cite{he2016deep}           & 28.8M                 & Inhouse endoscopic videos \\
GastroNet\_ViTS \cite{boers2024foundation}          & ViT Small \cite{dosovitskiy2020image}            & 23.5M                 & Inhouse endoscopic videos \\
EndoFM \cite{wang2023foundation}                   & ViT Base \cite{dosovitskiy2020image}            & 123.9M                  & Large-scale surgical dataset \\
GSViT \cite{schmidgall2024general}                    & ViT Base \cite{dosovitskiy2020image}             & 89.2M                 & Large-scale surgical dataset \\
EndoViT \cite{batic2024endovit}                  & ViT variant                 & 13.9M                 & Large-scale surgical dataset \\
\bottomrule
\end{tabular}
\begin{tablenotes}
    \scriptsize
    \item[a] \textsuperscript{*} All models were evaluated on both datasets except SurgNet\_ConvNeXTV2, ConvNeXTV2, and Timesformer, which were not evaluated on AutoLaparo.
\end{tablenotes}
\label{tab:model_overview}
\end{table} To establish ground truth for surgical phase recognition, we follow the two-step training strategy introduced by TeCNO \cite{czempiel2020tecno}. In the first step, we train a backbone network to predict phases from individual frames. In the second step, we apply a Multi-Stage Temporal Convolutional Network (MS-TCN) to refine these predictions by incorporating temporal context.

We train all models on an NVIDIA GeForce RTX 2080 Ti GPU using consistent training parameters across all experiments. In the first stage of training, we optimize the CAFormer-based feature extractor using a learning rate of 1e$^{-5}$ and cross-entropy loss. In the second stage, we train the MS-TCN on the extracted features using a learning rate of 7e$^{-4}$ for 200 epochs, again using cross-entropy loss. The resulting test accuracy is used as the ground truth score $G_m$ for model $\Phi_m$. This way, we obtain a set of scores $\{G_m\}_{m=1}^M$ as the ground truth to evaluate our pre-trained model rankings.

To compute our metrics, we first perform a single forward pass of the pre-trained model through all target examples to extract their features. Following the previous works~\cite{SFDA,logme,NLEEP}, we use weighted Kendall's tau $\tau$~\cite{10.1145/2736277.2741088} to evaluate the effectiveness of transferability metrics. Kendall's tau $\tau$ returns the ratio of concordant pairs minus discordant pairs when enumerating all $(_2^M)$ pairs of $\{T_m\}_{m=1}^M$ and $\{G_m\}_{m=1}^M$ as given by:
\begin{equation}
    \tau = \frac{2}{M(M-1)}\sum_{1\leq i \leq j M} \mathrm{sgn}(G_i-G_j)\mathrm{sgn}(T_i-T_j)
\end{equation}

Here $\mathrm{sgn}(x)$ is the signum function returning 1 if $x>0$ and -1 otherwise. In the weighted version of Kendall's tau $\tau$, the ranking performance of top-performing models is measured to evaluate transferability metrics. In principle, a higher $\tau$ indicates that the transferability metric produces a better ranking for pre-trained models. We evaluate LogME \cite{logme}, H-Score \cite{Hscore}, and TransRate \cite{transrate2}, corresponding respectively to evidence-maximization, information theory, and stability-based approaches.

\section{Results and Discussion}

\subsection{Overall Results}

\begin{figure*}[h]
  \centering
  \begin{subfigure}[b]{0.49\textwidth}
    \includegraphics[width=\textwidth]{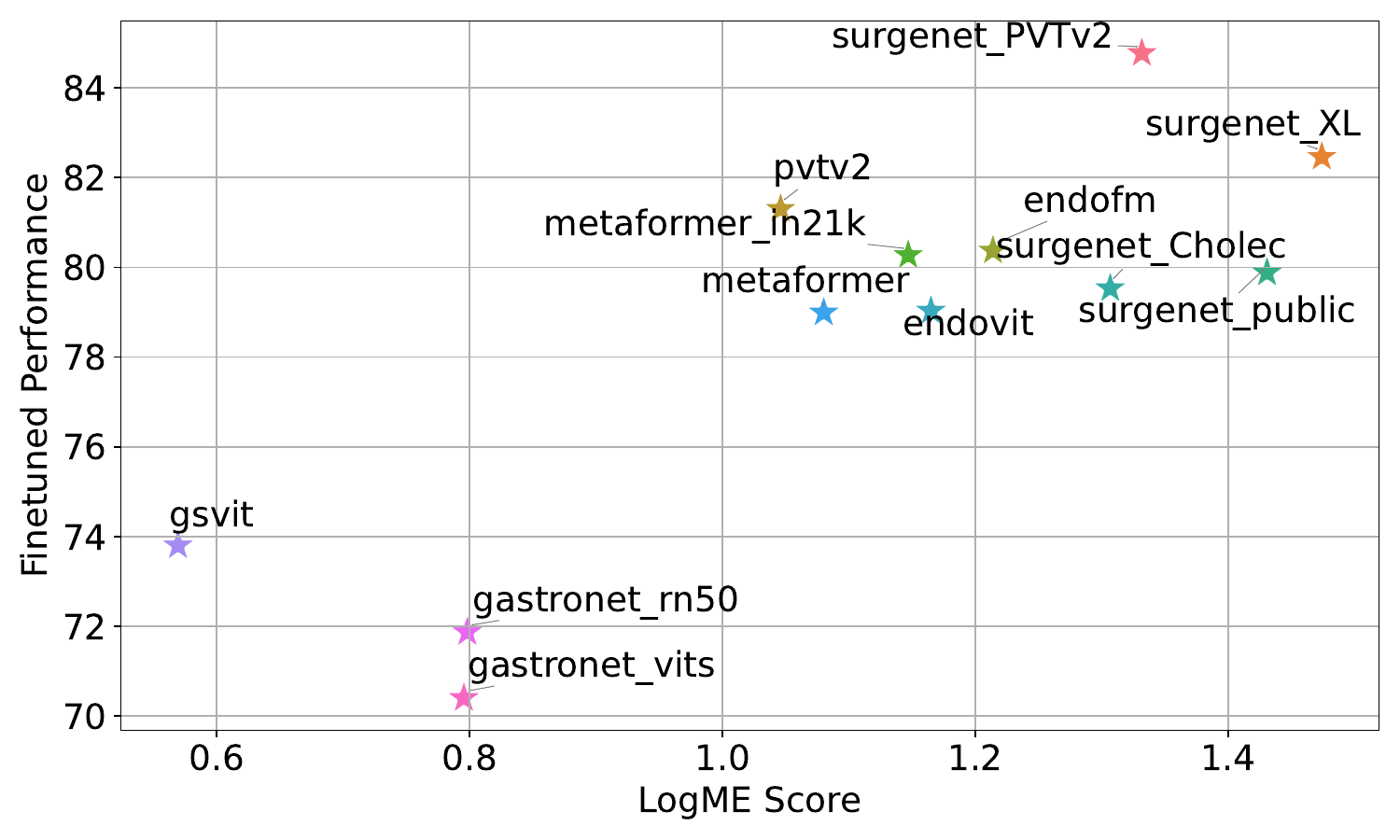}
    \caption{LogME vs Accuracy for AutoLapro}
    \label{fig:subfigure1}
  \end{subfigure}
  \hfill
  \begin{subfigure}[b]{0.49\textwidth}
    \includegraphics[width=\textwidth]{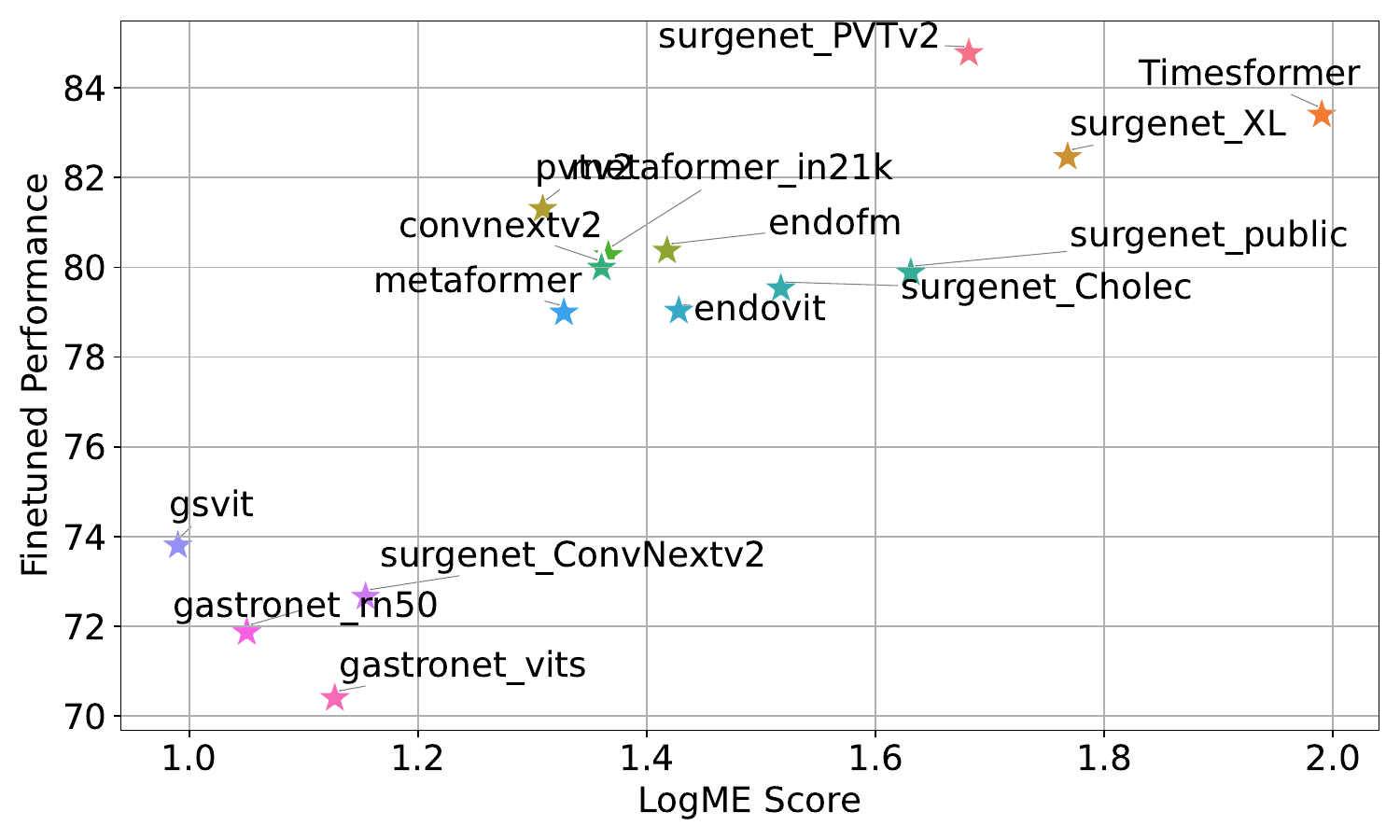}
    \caption{LogME vs Accuracy for RAMIE}
    \label{fig:subfigure2}
  \end{subfigure}
  \caption{Relationship between mean‐aggregated LogME transferability score and downstream accuracy for (a) AutoLaparo and (b) RAMIE.}
  \label{fig:main_figure}
\end{figure*}

Table \ref{tab:performances} summarizes our experimental findings. LogME (mean) achieves Pearson’s \(r=0.627\) on AutoLaparo and \(r=0.653\) on RAMIE, with Kendall’s \(\tau=0.833\) and \(\tau=0.835\), respectively; LogME (min) yields slightly higher Pearson correlations (\(r=0.655, 0.674\)) and matching \(\tau\) values (\(0.825, 0.835\)), while LogME (max) falls behind (\(r\le0.608, \tau\le0.747\)).\begin{table}[h!]
    \centering
    \caption{Performance of each metric with different summary functions on RAMIE and AutoLaparo, whereby $r$ denotes Pearson and $\tau$ denotes Kendall correlation. }
\begin{tabular}{lcccc}
\toprule
 Metric & AutoLaparo ($r$) & AutoLaparo ($\tau)$ & RAMIE ($r$) & RAMIE ($\tau$) \\
\midrule
LogME (mean) & 0.627 & 0.833 & 0.653 & 0.835 \\
LogME (min) & 0.655 & 0.825 & 0.674 & 0.835 \\
LogME (max) & 0.594 & 0.696 & 0.608 & 0.747 \\
\midrule
Hscore (mean) & 0.056 & -0.203 & -0.024 & 0.365 \\
Hscore (min) & 0.501 & 0.158 & 0.056 & 0.352 \\
Hscore (max) & -0.031 & -0.190 & 0.074 & -0.138 \\
\midrule
TransRate (mean) & -0.252 & -0.195 & -0.095 & -0.181 \\
TransRate (min) & -0.194 & -0.188 & -0.052 & -0.201 \\
TransRate (max) & -0.208 & -0.203 & -0.095 & -0.187 \\
\bottomrule
\end{tabular}
    \label{tab:performances}
\end{table} H-Score (min) reaches modest positive alignments (\(r=0.501, 0.056\); \(\tau=0.158, 0.352\)), but its other variants remain near zero or negative, and all TransRate configurations produce negative \(\tau\) (\(\approx-0.19\)), indicating an inverted ranking relative to true performance. Figure \ref{fig:main_figure} plots the mean‐aggregated LogME scores against fine‐tuning accuracy on both datasets, illustrating the strong linear trend that underlies these Pearson and Kendall correlations.

Overall, LogME outperforms H-Score and TransRate (\(\tau\approx0.83\) on both datasets), while TransRate scores are negatively correlated with actual performance, ranking stronger models lower.
 The up to 0.14 boost in $\tau$ from max to min aggregation highlights the critical role of summary choice, and the divergence between Pearson’s $r$ and Kendall’s $\tau$ for H-Score(max) underscores the need to report both linear and rank correlations.
\subsection{Ablation Studies}
To assess how much LogME’s ranking relies on performance extremes, we iteratively pruned our models and tracked the change in weighted Kendall’s $\tau_w$.  

\begin{itemize}
  \item \textbf{RAMIE dataset.} Starting from $\tau_w=0.835$ (full pool; Table~\ref{tab:performances}), removing the three top performers (TimeSformer, SurgNet$\_$PVTv2, SurgNet$\_$XL) causes $\tau_w$ to drop to 0.24. Further removing the three weakest models inverts the correlation to $\tau_w=-0.07$. These steep declines demonstrate that LogME’s discriminative power hinges on a wide span of model qualities—once you eliminate both high and low extremes, little ranking signal remains.
  \item \textbf{AutoLaparo dataset.} From $\tau_w=0.825$, excluding just the single strongest model (SurgNet$\_$XL) reduces $\tau_w$ to 0.36, underscoring LogME’s sensitivity to influential outliers.
\end{itemize}
Based on our experiments and ablations, SITE via LogME excels when candidate models span a wide performance range, clearly separating strong from weak, but its ranking degrades or even inverts when models cluster within a narrow (5–7 $\%$) band. H-Score delivers only modest signal under minimum aggregation, and TransRate consistently reverses true ordering. Practitioners should therefore ensure sufficient diversity in their model pool or complement LogME with additional evaluation criteria.

Our current evaluation is limited to two surgical-phase datasets and three transferability metrics, and lacks formal thresholds for judging small score differences (e.g., when a 1 $\%$ gap is practically meaningful), which may hinder fine-grained model selection. Nonetheless, SITE methods demonstrate strong potential to predict downstream performance without costly fine-tuning by assessing the compatibility between pre-trained embeddings or predictions and target labels. This enables efficient, objective model ranking and reduces subjective bias in surgical video analysis. 




\section{Conclusion}
In this work, we introduced a source‐independent transferability estimation (SITE) framework for surgical phase recognition and conducted a systematic evaluation of three transferability metrics across two surgical video datasets. Our study shows that while evidence‐maximization approaches like LogME deliver reliable model rankings, information‐theoretic and stability‐based metrics offer limited or even misleading signals in this setting. An ablation analysis highlights the importance of maintaining a broad performance range among candidate models or augmenting SITE with additional validation when models are closely clustered. In summary, we (i) formalize SITE for surgical phase recognition, (ii) benchmark three state-of-the-art transferability metrics on two datasets, (iii) perform ablations to expose their sensitivity to model diversity, and (iv) derive practical recommendations for practitioners.

Future work will broaden SITE evaluation by incorporating additional surgical and non‐surgical video datasets across modalities and annotation protocols, and by developing new transferability metrics tailored to video data (e.g.\ temporal embeddings, graph‐based similarity). We will integrate statistical significance testing (bootstrap CIs, hypothesis tests) to identify practically meaningful score gaps, and investigate the theoretical basis of transferability in terms of feature‐space geometry, task similarity, and domain shift. Finally, we plan to build interactive tools that combine SITE metrics, score visualizations, and uncertainty estimates to guide model selection without exhaustive fine‐tuning.
\bibliographystyle{splncs04}
%
\bibliography{main}

\end{document}